\newcommand{\beq}{\begin{equation}}
\newcommand{\eeq}{\end{equation}}
\newcommand{\beqa}{\begin{eqnarray}}
\newcommand{\eeqa}{\end{eqnarray}}
\newcommand{\ket}[1]{| #1 \rangle}
\newcommand{\bra}[1]{\langle #1 |}

\documentclass[pra,twocolumn,showpacs]{revtex4}

\begin{document}


\title{Single-particle nonlocality and entanglement with the vacuum}

\author{Gunnar Bj\"{o}rk}
\thanks{Presently on leave at Departamento de \'{O}ptica, 
Facultad de Ciencias F\'{\i}sicas, 
Universidad Complutense, 28040 Madrid, Spain}
\email{gunnarb@ele.kth.se}
\homepage{http://www.ele.kth.se/QEO/}
\affiliation{Department of Microelectronics and 
Information Technology, Royal Institute of Technology 
(KTH), Electrum 229, SE-164 40 Kista, Sweden}
\author{Per Jonsson}
\affiliation{Department of Microelectronics and 
Information Technology,  Royal Institute of Technology 
(KTH), Electrum 229,  SE-164 40 Kista, Sweden}
\author{Luis L. S\'{a}nchez Soto}
\affiliation{Departamento de \'{O}ptica, 
Facultad de Ciencias F\'{\i}sicas, 
Universidad Complutense, 28040 Madrid, Spain}

\date{\today}

\begin{abstract}
We propose a single-particle experiment that is equivalent 
to the conventional two-particle experiment used to demonstrate 
a violation of Bell's inequalities. Hence, we argue that quantum 
mechanical nonlocality can be demonstrated by single-particle states. 
The validity of such a claim has been discussed in the literature, but 
without reaching a clear consensus. We show that the disagreement 
can be traced to what part of the total state of the experiment one assigns 
to the (macroscopic) measurement apparatus. However, with a 
conventional and legitimate interpretation of the measurement process 
one is led to the conclusion that even a single particle can show nonlocal 
properties.
\end{abstract}


\maketitle

\section{Introduction}

Single-photon sources are coming of age. The most common way 
to produce single-photon states with random emission times is 
to use photon-pair emission in spontaneous parametric 
down-conversion~\cite{Hong,Rarity,Kwiat}, followed by 
detection of one of the emitted photons. However, recently 
sources able to deliver a single photon \textit{on demand} 
have been  demonstrated, such as single-molecule 
emitters~\cite{De Martini,Lounis,Brunel,Fleury}, 
electrically-driven semiconductor p-i-n junctions~\cite{Kim}, 
color  centers in diamond~\cite{Kurtsiefer,Brouri}, and 
semiconductor  quantum dots~\cite{Michler,Michler 2,Santori,Zwiller}. 
As deterministic single-photon sources are being refined, it is 
relevant to discuss their potential in quantum information applications 
and in fundamental tests of physics. In this work we focus on 
the second of these questions, and specifically address if and 
how a single photon can be used to demonstrate quantum nonlocality. 

Nonlocal properties of a single particle have been discussed by several 
authors~\cite{Tan,Hardy,Czachor,Revzen,Peres,Home,Gerry,MMichler}. 
However, most of the proposals have been (or can be) criticized for 
various reasons. The proposals by Tan, Walls, and Collett~\cite{Tan} 
and by  Hardy~\cite{Hardy} have been criticized as being multiparticle 
demonstrations of nonlocality in disguise~\cite{Greenberger,Greenberger 2} 
and for other reasons~\cite{Santos,Vaidman,Hardy 2}. The criticism has 
been refuted as partially being a ``semantic issue", pertaining to 
the interpretation of the meaning of ``single-particle nonlocality"~\cite{Hardy 2}. 
The proposals put forth by Czachor~\cite{Czachor} and by Home and 
Agarwal~\cite{Home} are based on Mach-Zehnder interferometers, 
so the measurement does not take place in two spacelike separated locations
and, consequently, the tests are not  ``loop-hole free".  The remarkable 
inequality   found by Revzen and Mann~\cite{Revzen} is not derived 
in terms of experimentally testable entities; instead it demonstrates that 
the statistical interpretation of quantum mechanics is at odds with a local 
hidden-variables theory. Peres~\cite{Peres} presents a very clear and 
concise discussion of single-particle nonlocality, but offers no suggestion 
how to experimentally implement the projective measurements his 
discussion centers around. In the experiment proposed by Gerry~\cite{Gerry} 
the nonlocal properties of a single photon are transferred to two atoms 
prior to the measurement and, therefore, what is finally measured 
are the correlations between two particles, which makes the claim of 
single-particle nonlocality somewhat weaker. To the best of our knowledge, 
only one experiment has hitherto been performed that claims to be a 
single-particle test of noncontextual hidden variables~\cite{MMichler},
but, unfortunately, it is also based on self-interference in a Mach-Zehnder 
interferometer. Thus, one can conclude that so far no loop-hole 
free demonstration of single-particle nonlocality has been performed. 

In this paper we shall re-examine single-photon nonlocality. Specifically, 
we shall point out why early, conceptually simple, proposals~\cite{Tan,Hardy} 
are very hard to implement in practice. We will argue that, while the proposal 
of Tan, Walls, and Collett~\cite{Tan} in our opinion is sound, by replacing the 
quadrature amplitude measurements the proposals are based on, with phase 
measurements, one will obtain a \textit{single-particle} equivalent to Bohm 
and Aharonov's two-particle version of the EPR gedanken experiment~\cite{Bohm}. 
The advantage with our proposal is that Bohm and Aharonov's version of the 
EPR experiment is well understood and familiar to most physicists. We 
will also show that an experiment involving relative phase, rather than phase, 
will be experimentally much simpler.

At first, it may seem counterintuitive that a single particle could have nonlocal 
properties, since observation of nonlocality would entail detection of some 
property of the particle at two spacelike separated locations. Clearly, detection 
of the particle at one location would immediate nullify any possibility to 
simultaneously record the particle, or any property associated with the particle, 
at another location. The resolution of this apparent conflict is provided by 
quantum-mechanical duality. Recall that any particle also has wavelike properties, 
and while the word ``particle" brings to mind a pointlike, localized entity, waves 
are usually  thought of as delocalized. Hence, the nonlocal properties of a single 
particle should naturally be sought in its wavelike properties. 

Going back to the particle viewpoint, in order to be able to simultaneously record 
some joint property of a particle at spacelike separated regions, the particle must 
be prepared in a superposition state of being localized at one or the other location. 
The only other state that we can invoke in the superposition is the vacuum and, 
hence, single-particle nonlocality entails entanglement with the vacuum. 

Entanglement with the vacuum is a controversial issue. We cite from 
Ref.~\cite{Greenberger 2}: ``We point out that it can be very misleading to 
discuss entangled states in Fock space. Some states that seem to be entangled 
there are merely single-particle states in configuration space, with no EPR-type 
nonlocality implications. Other states that seem to be product states are clearly 
entangled states in configuration space. Especially misleading is the concept of 
`states entangled with the vacuum'." We do not share the opinions voiced by 
Greenberger, Horne, and Zeilinger: since the Fock basis is a complete basis, 
it is just as good as any other to express and calculate quantum physics, 
provided the configuration space, or the modes of  the system, are unambiguously 
defined.  Along this paper we shall show that this concept of ``entanglement
with the vacuum", when properly used, could be very useful and, with a 
conventional interpretation of the measurement process, lead to the 
conclusion of the nonlocality of a single particle.

\section{A single-particle Bell-inequality violation}

The smallest state space in which it is possible to demonstrate EPR 
effects is a four-dimensional space (the product space of two spacelike 
separated two-state spaces). The basis vectors of this four-dimensional 
Hilbert space is conventionally taken as the Bell basis. In this paper 
we will focus on the Bell state
\beq
\ket{\Psi_-}= \frac{1}{\sqrt{2}} (\ket{1}_a \otimes \ket{0}_b - 
\ket{0}_a \otimes \ket{1}_b) ,
\label{eq: Bell plus}
\eeq
where the indices $a$ and $b$ usually are taken to refer to ``particle $a$"
and ``particle $b$", and where
\beq
{}_a \langle 0 \ket{1}_a = {}_b \langle 0 \ket{1}_b = 0  .
\label{eq: orthogonality}
\eeq 
We stress that sufficient requirements for $\ket{\Psi_-}$ to display nonlocal 
properties is that the indices $a$ and $b$ represent modes, or configurations, 
that are spacelike separated and that the orthogonality condition 
(\ref{eq: orthogonality}) is satisfied. What physical states the kets 
$\ket{0}_a$, $\ket{1}_a$, $\ket{0}_b$, and $\ket{1}_b$  represent 
is \textit{irrelevant} from a strictly fundamental point of view.

A single-particle state of the form (\ref{eq: Bell plus}) is  
\beq
\ket{\Psi}= \frac{1}{\sqrt{2}} (\ket{1,0} - \ket{0,1})  ,
\label{eq: vacuum entanglement}
\eeq
where we have suppressed the indices, abbreviated $\ket{m}\otimes
\ket{n}$ to $\ket{m,n}$, and used the number basis. This is a 
single particle entangled with the vacuum (in the following 
we shall assume that the particle is a photon). The state can be 
generated by letting a one photon state prepared in a well specified 
spatio-temporal mode impinge on a 50:50 beam splitter, 
see Fig.~\ref{fig: simplest setup}. If the beam splitter is oriented 
so that the transmitted and reflected modes propagate perpendicularly, 
the two emerging wave packets will be separated by a spacelike 
distance, and a loop-hole free Bell test could in principle be performed. 
The state, after propagation during times $\tau_c$ and $\tau_d$ along 
the two ``arms" $c$ and $d$, will become
\beq
\ket{\Psi (\tau_c, \tau_d)} = 
\frac{1}{\sqrt{2}} ( e^{i \omega \tau_c} \ket{1,0} - 
e^{i \omega \tau_d} \ket{0,1}) ,
\label{eq: vacuum entanglement t}
\eeq
where $\omega$ is the angular frequency of the light. 

Let us now discuss how, in principle, the nonlocal properties of this 
state, identical in form to $\ket{\Psi_-}$, could be measured. To this 
end, let us consider the projectors
\beqa
\frac{1}{2}  [ 
e^{i (\phi_c +\omega \tau_c)} \ket{1} + \ket{0} ] \otimes 
[ e^{-i (\phi_c +\omega \tau_c)} \bra{1} + \bra{0}  ] 
\label{eq: projector 3}   \\
& & \nonumber \\
\frac{1}{2}  [ 
e^{i (\phi_d + \omega \tau_d)} \ket{1} + \ket{0} ] \otimes 
[ e^{-i (\phi_d + \omega \tau_d)} \bra{1} + \bra{0} ]
\label{eq: projector 4}
\eeqa
acting on the state in arms $c$ and $d$, respectively. These 
projectors are Pegg-Barnett phase projectors~\cite{Pegg} in a 
two-dimensional Hilbert space. Calculating the associated projection 
probabilities of the state $\ket{\Psi(\tau_c, \tau_d)}$ one finds that 
\beq
P(\phi_c) = P(\phi_d) = \frac{1}{2}, \quad 
P(\phi_c, \phi_d) =\frac{1}{2} \sin^2[(\phi_c - \phi_d)/2] , 
\label{eq: probabilities}
\eeq
where, e. g., $P(\phi_c)$ denotes the probability to detect the phase 
$\phi_c$ in arm $c$, and $P(\phi_c, \phi_d)$ denotes the joint probability 
to detect the phase $\phi_c$ in arm $c$ and the phase $\phi_d$ in arm $d$. 
The probabilities are \textit{identical} to those encountered in Bohm and 
Aharonov's version of the EPR paradox~\cite{Bohm}, and in Bell subsequent 
analysis of bounds on local hidden variables and quantum predictions~\cite{Bell}. 
This is simply because the state $\ket{\Psi}$ is identical in form to the Bell 
state $\ket{\Psi_-}$.

There is nothing in quantum theory ruling out an experimental 
implementation of the projectors (\ref{eq: projector 3}) and 
(\ref{eq: projector 4}) as classical measurement devices. Consequently, 
we have to assume that classical measurement devices exist that 
can implement them, just like we assume that there exist classical 
devices implementing, e.g., the projector $\ket{1}\bra{1}$. Therefore
, one is lead to the conclusion that quantum theory is nonlocal 
even for single particles. However, the sensitive time dependence 
of (\ref{eq: projector 3}) and (\ref{eq: projector 4}) implies that an 
experimental configuration would have to be stable in time by a 
fraction of an optical period, and consequently in space by a fraction 
of a wavelength. Since $\ket{\Psi (\tau_c, \tau_d)}$ is not an 
eigenstate of the free-space propagation Hamiltonian, 
it will be difficult to implement such an experiment. This is 
the main reason why the experiments proposed in Refs.~\cite{Tan} 
and \cite{Hardy} have not yet been attempted. The conventional 
(and experimentally simpler) tests of Bell's inequalities are based on 
two-particle states that are eigenstates of the free-space propagation 
Hamiltonian. We stress that going from single-particle to two-particle 
nonlocality tests simplifies things tremendously from an experimental 
point of view, but \textit{changes nothing} from a fundamental 
point of view.

Peres~\cite{Peres} has pointed out that the projectors (\ref{eq: projector 3}) 
and (\ref{eq: projector 4}) do not commute with the total photon-number 
operator. He concludes that ``nonlocal effects may thus appear for 
an initial state that contains a single particle, provided that the final 
state may contain two." While the statement is correct, it should 
be an unsatisfactory answer to whose who argue that single-particle states 
cannot be used to demonstrate nonlocality. The reason is that while the 
final (post-measurement) state will contain two photons with probability 1/4, 
it will contain \textit{no photon} with the same probability. If the first 
outcome is taken as an argument that our, and earlier, similar proposal are 
only a demonstration of multiphoton nonlocality in disguise, the same logic leads 
to the conclusion that the second outcome indicates that nonlocality can also 
be demonstrated with no particles. However, both ``conclusions" are equally 
misleading, since they ``explain" the nonlocal characteristics of the 
pre-measurement state in terms of probabilities derived from the 
post-measurement state. 

The pre-measurement state $\ket{\Psi}$ is fully characterized by 
two binary, truth propositions~\cite{Zeilinger}. Expressed operationally, 
they are: (a) the sum of the photon numbers measured in the two arms 
is unity; and (b) the phase measured in one arm will always differ 
by $\pi$ from the phase measured in the other arm. Since the phases 
measured at the two locations contain an element of reality (the 
$\pi$ difference is certain) our proposed experiment avoids ``the law 
of the excluded muddle"~\cite{Bernstein}. The title of our paper is 
simply a summary of the consequences of truth propositions (a) and (b).

\section{Single-photon nonlocality based on relative phase}

Now let us return to the experiment. The difficulties associated with 
single-particle nonlocality based on measurement of phase can be 
overcome by measuring relative phase instead of phase. To this 
end, consider the schematic setup depicted in Fig.~\ref{fig: setup}. 
Incident on a polarizing beam splitter is a product state between a 
single-photon state and a coherent state with a mean photon number 
$2 |\alpha|^2$ (for simplicity, and without loss of generality, we shall 
assume that $\alpha$ is real). The states are both linearly polarized at 
a direction 45 degrees from the horizontal (in the following, vertical and 
horizontal polarizations will be denoted V and H, respectively). Expressing 
the state in a vertical-horizontal linear-polarization four-mode basis, the 
impinging state can be written as
\beq
\frac{1}{\sqrt{2}} ( \ket{1,0} - \ket{0,1}) 
\otimes \ket{\alpha,\alpha},
\label{eq: HV basis}
\eeq
where we take the kets (left to right) to denote the modes 
$a$V, $a$H, $b$V, and $b$H. After the polarizing beam splitter, 
the state becomes
\beq
\frac{1}{\sqrt{2}} ( \ket{1,\alpha,\alpha,0} - \ket{0,\alpha,\alpha,1})
\label{eq: PBS state}
\eeq
if expressed in the modes $c$V, $c$H, $d$V, and $d$H. In absence 
of polarization dispersion (assume, e. g., that the wave packets propagate 
in vacuum or air), the state after mode $c$ has evolved during time $\tau_c$ 
and mode $d$ during the time $\tau_d$ will be
\beq
\frac{1}{\sqrt{2}} ( e^{i \omega \tau_c}
\ket{1, e^{i \omega \tau_c} \alpha, e^{i \omega \tau_d} \alpha, 0} - 
e^{i \omega \tau_d} \ket{0, e^{i \omega \tau_c} \alpha, 
e^{i \omega \tau_d} \alpha, 1}) .
\label{eq: PBS state 2}
\eeq
The exponential phase factors in Eq.~(\ref{eq: PBS state 2}) 
preserve the relative phase between the modes in each arm. 
In absence of birefringence the relative phase between the two modes 
is a constant of motion. 

To measure the relative phase between the states in arm $c$ we 
introduce, in each two-mode energy manifold $n>0$, the projector 
with eigenstate 
\beq
\ket{\xi^{(n)}(\phi)} = \frac{1}{\sqrt{1 + n/\alpha^2}} 
\left [ \frac{\sqrt{n}}{\alpha} \ket{0,n} + 
e^{i \phi}\ket{1,n-1} \right ] . 
\label{eq: projector 3B}
\eeq
Note that this projector is time-independent, and it is therefore 
also invariant under translations along the arm. In the $n=0$ 
energy manifold, there is only one associated state, so in this 
manifold there exist no relative-phase-dependent projector. 
In all other manifolds the eigenstate $\ket{\xi^{(n)}(\phi)}$ 
is similar in form to the eigenstates of the relative-phase 
operator~\cite{Luis}. The projection probabilities in 
the two arms on the states $\ket{\xi^{(n_c)}(\phi_c)}$ and 
$\ket{\xi^{(n_d)}(\phi_d)}$, respectively, and the joint probability 
of detecting the relative phases $\phi_c$ and $\phi_d$ and the 
photon numbers $n_c>0$ and $n_d>0$ become:
\begin{eqnarray}
P(n_c, \phi_c) & = & \frac{e^{\alpha^2} 
\alpha^{2(n_c-1)}}{(1+\ n_c / \alpha^2)(n_c-1)!} ,
\nonumber \\
P(n_d, \phi_d) & = & \frac{e^{\alpha^2} 
\alpha^{2(n_d - 1)}}{(1+ n_d / \alpha^2)(n_d-1)!} ,  \\
P(n_c, n_d, \phi_c, \phi_d) & = & 2 
P(n_c, \phi_c) P(n_d, \phi_d) \sin^2[(\phi_c- \phi_d)/2 ] .
\nonumber 
\label{eq: n manifold}
\end{eqnarray}
The relative-phase probabilities $P(n_c, \phi_c)$ and 
$P (n_d, \phi_d)$ are independent of the settings of $\phi_c$ 
and $\phi_d$. Summing, e. g., $P(n_c, \phi_c)$ over $n_c$, 
one finds that probability of obtaining the relative phase $\phi_c$ 
approaches 1/2 as the coherent state excitation increases. In 
Fig.~\ref{fig: converge} we have plotted the difference $1/2 - 
\sum_{n_c=1}^\infty P(n_c, \phi_c)$ as a function of the mean 
photon number $\alpha^2$ of the coherent state. In Fig.~\ref{fig: cos} 
we show how closely the projector defined by summing 
$P(n_c, n_d, \phi_c, \phi_d)$ over $n_c$ and $n_d$ approximates 
the ideal projector for $\alpha^2=3$ and $\alpha^2=10$. To quantify 
the deviation between the joint relative-phase probability and the joint 
phase-projector probability $P(\phi_c, \phi_d)$, we have also plotted 
the maximum difference between the two (that is, for $\phi_c - 
\phi_d = \pm \pi$) in Fig.~\ref{fig: converge}.

Before returning to the central question of the paper, namely 
nonlocal properties of single particles, let us briefly discuss some 
technical aspects of our relative-phase proposal. One way to 
experimentally implement the proposal would be to make devices 
that, in each arm and each manifold $n>0$, perform the transformation
\beq
\ket{m,n-m} \bra{\xi^{(n)}(0)} ,
\label{eq: unitary}
\eeq
where $0<m \leq n$ (in this context it is irrelevant how all states orthogonal 
to $\ket{\xi^{(n)}(0)}$ are transformed). In this way, detection of the state 
$\ket{\xi^{(n)}(0)}$ is converted to the much simpler (photon counting) detection 
of the state $\ket{m,n-m}$. In manifold $n=1$ and $n=2$ such transformations 
have be accomplished by the means of linear components, i.e., beam splitters and 
phase plates~\cite{Grangier,Trifonov}. To make the projectors depend in 
the desired way on the relative phases $\phi_c$ and $\phi_d$, variable 
birefringence components, such as birefringent liquid crystal cells, or birefringent 
wedges, could be inserted in the arms prior to the projective measurements~\cite{Trifonov}. 
In manifolds $n>2$ the transformations will require a nonlinear Hamiltonian~\cite{Bjork}, 
of the same level of technical difficulty as implementing a quantum optical controlled 
NOT gate. Hence, our proposal is experimentally challenging at the moment, but 
is serves to demonstrate that single particles can indeed be used to show 
Bell-type correlations.

Now, let us go back to the interpretation of the proposed experiment. 
Clearly the measurement involves more than a single particle. However, 
all the nonlocal properties demonstrated by such an experiment are carried 
by a single particle. Hardy~\cite{Hardy 2} suggested four criteria for 
unambiguous demonstration of single-photon nonlocality. Slightly abbreviated 
they are: (I) There should be a single-photon source and two quantum 
channels leading to spacelike separated measurement regions. 
In addition there may be classical channels between the measurement 
regions carrying classical information. (II) Photon detectors placed directly 
into the quantum channels will detect no more that one photon in the measured 
spatio-temporal modes. (III) If any of the quantum channels are blocked, no 
violation of locality can be observed. (IV) The results of the experiment violate 
locality. Our proposal meets all four criteria. The coherent state is a classical 
phase reference, copropagating with the single photon only to make the 
experiment less sensitive to measurement imperfections due to limited 
precision, or nonfundamental noise, in the time and space coordinates. 

In principle, the coherent states could be produced locally. Since the phase 
stability of a laser are fundamentally limited only by the cold-cavity 
linewidth of the resonator and the energy stored in the cavity, there i
s no fundamental limit for how long two lasers can stay in 
synchronism~\cite{Schawlow}. Hence, \textit{in principle}, two 
lasers could be adjusted (by a homodyne measurement) so that 
their respective phases coincided, then transported to two remote 
locations. Within a time proportional to the inverse linewidths, the lasers 
would stay synchronized and the experiment could be performed 
without a classical communication channel, that is, as a
``black-box measurement". Hence, the needed phase reference 
provided by the coherent states should be interpreted as internal 
states associated with the macroscopic measurement apparata 
implementing the projectors (\ref{eq: projector 3}) and  (\ref{eq: projector 4}). 
That such an interpretation is both customary and legitimate  has already 
been argued by Peres~\cite{Peres,Peres 2}. Enclosing the lasers in 
``black boxes", such an experiment performed on a single photon in the 
state $\ket{\Psi}$ could yield identical measurement statistics and a 
similar measurement configuration as spin analysis of two spin 1/2 
particles in a singlet state.

In a more realistic scenario one can envision two individual (slave) 
lasers, one at the end of each arm, that are regularly synchronized 
by a short pulse of light from a centrally placed master laser, see 
Fig.~\ref{fig: black box setup}. During a short time, the classical 
channels are used to synchronize the slave lasers to the master laser 
and no measurements are done. Then the master laser is switched off 
and a series of single-photon relative-phase measurements, where the 
coherent states are produced locally, are made. The process is then 
repeated with a frequency higher than the lasers linewidth (that is 
assumed to impose a more stringent requirement on the repetition 
frequency than the mechanical drift and vibrations of the experimental 
setup). This is essentially how ``clock recovery" is performed in 
coherent optical communications systems. In these, local 
oscillator synchronization (albeit usually at much lower 
frequencies) is an integral part of the whole system. We see 
no fundamental reason why such a synchronization scheme 
could not be implemented at optical frequencies. In fact, 
more than ten years ago, such schemes were seriously being 
discussed in the context of coherent optical communication~\cite{Kikuchi}.  

\section{Summary}

In conclusion, we have proposed an experiment that demonstrates 
nonlocal properties of a single particle. Our proposal is a single-particle 
analog to the two spin 1/2 particle experiment proposed by Bohm and 
Aharonov. Our experiment will be difficult to perform in practice, 
because most, if not all, detectors with a single-quanta sensitivity 
are particle (or energy) counters. Since single-particle nonlocality 
must not involve direct particle detection, but phase, or relative-phase 
detection, a unitary transformation must be used to convert these 
properties to properties measurable by a particle counter. 

We have argued that the states providing the needed phase references 
should be ascribed to the macroscopic meter. At any rate these states 
carry no nonlocal characteristics. The inevitable conclusion must 
be that any spacelike separated state fulfilling~(\ref{eq: Bell plus}) 
and (\ref{eq: orthogonality}) has nonlocal properties, independent of 
what the kets represent. It is our hope that within a relatively near 
future the experiment we have proposed will be implemented experimentally.

\begin{acknowledgments}
We thank Prof. D. Greenberger, Prof. A. Zeilinger, and Prof. F. De Martini 
for making Ref.~\cite{Greenberger 2} available to us. This work was 
supported by the Swedish Foundation for Strategic Research (SSF), 
the European Union through program IST-1999-10243 (S4P), and 
L M Ericssons stiftelse f\"{o}r fr\"{a}mjande av elektroteknisk forskning. 
\end{acknowledgments}


\newpage

\begin{figure}
\caption{A single photon is split in a spacelike fashion by a 50:50 beam splitter. 
At the end of two arms, $c$ and $d$, the wave packets' phase projections at 
angles $\phi_c$ and $\phi_d$ are measured at times $\tau_c$ and $\tau_d$ after 
the single photon was generated.}
\label{fig: simplest setup}
\end{figure}

\begin{figure}
\caption{A single-photon wave packet and a coherent state (prepared in 
matching spatio-temporal modes) impinge toward two ports of a polarizing 
beam splitter (PBS). The states are linearly polarized at 45 degrees from 
the horizontal. At the end of the arms $c$ and $d$ the probabilities of the 
relative phases $\phi_c$ and $\phi_d$ of the spacelike separated state are 
measured.}
\label{fig: setup}
\end{figure}

\begin{figure}
\caption{The lower curve shows the difference 
$1/2-\sum_{n_c=1}^\infty P(n_c, \phi_c)$ as a function of the coherent 
state excitation $\alpha^2$. The probability $P(n_c, \phi_c)$ is independent 
of $\phi_c$. Symmetry implies that the same relations hold for $P(n_d, \phi_d)$. 
The upper curve shows the maximum deviation (occurring for $\phi_c - \phi_d=\pm \pi$) 
between the desired joint probability $P(\phi_c, \phi_d)$ and the joint probability 
$\sum_{n_c=1}^\infty \sum_{n_d=1}^\infty P(n_c, n_d, \phi_c, \phi_d)$.}
\label{fig: converge}
\end{figure}

\begin{figure}
\caption{The joint probability $P(n_c, n_d, \phi_c, \phi_d)$ to detect the 
two relative phases $\phi_c$ and $\phi_d$ as a function of the difference 
between them if the average excitation of the coherent state is $\alpha^2=3
$ (dashed line) and $\alpha^2=10$ (solid line). The dotted line shows the 
desired form of the joint probability.}
\label{fig: cos}
\end{figure}

\begin{figure}
\caption{A relative-phase measurement setup where the phase references are 
produced locally. Only one such ``black-box" relative-phase meter is shown in 
full detail. The slave laser phase is regularly synchronized to the master laser 
phase using, e.g., a Pound-Drever servo loop. During a time short compared to 
the slave laser's inverse linewidth, the relative phases can be measured locally 
in arm $c$ and $d$ without need for any master laser signal. The synchronization 
and measurement cycle can subsequently be repeated.}
\label{fig: black box setup}
\end{figure}

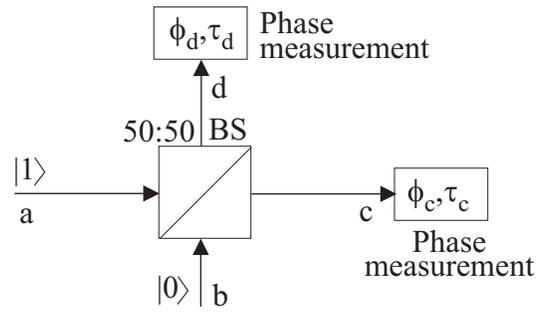
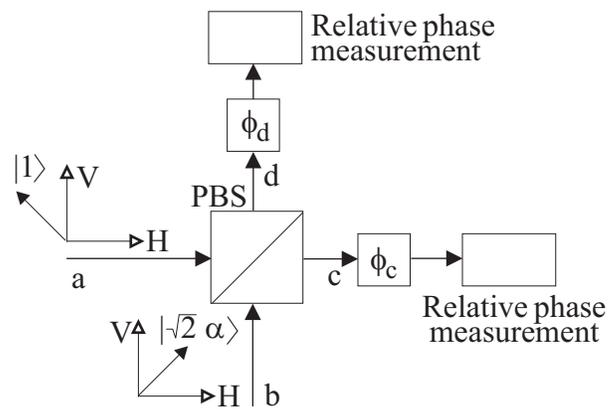
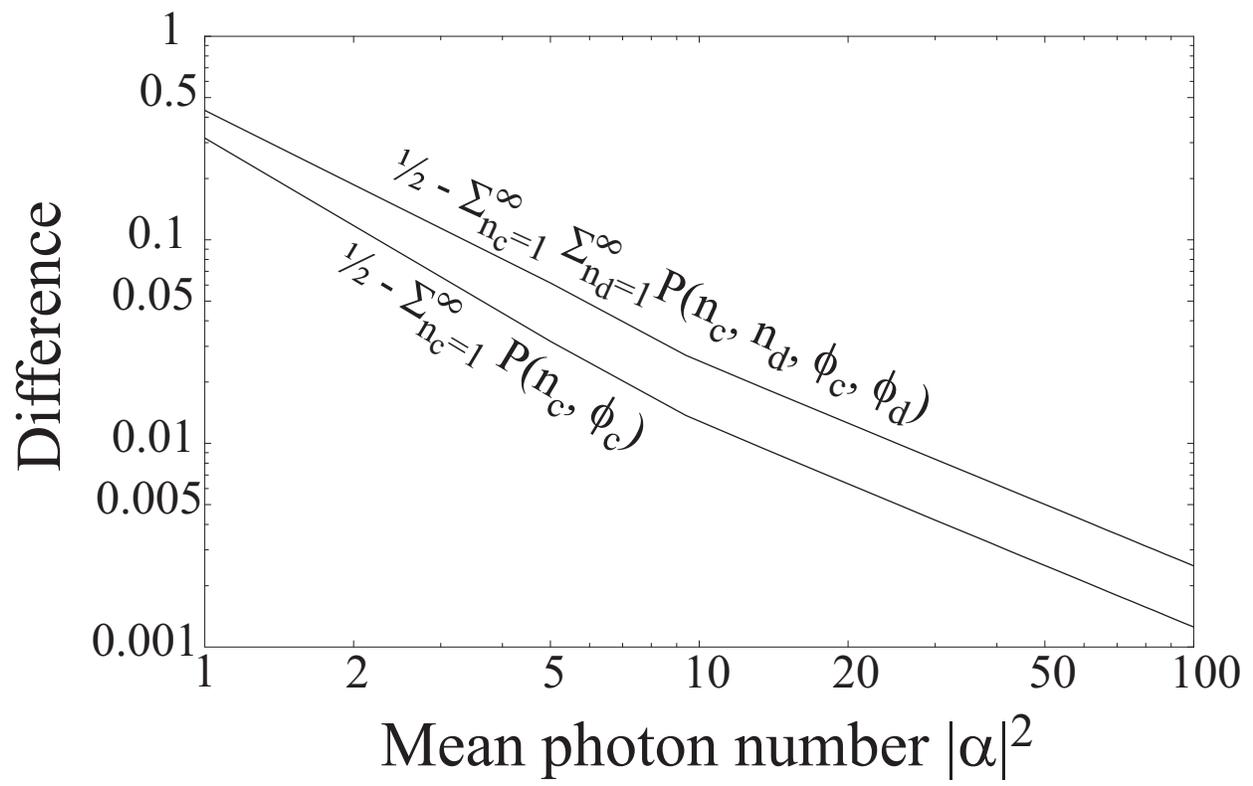
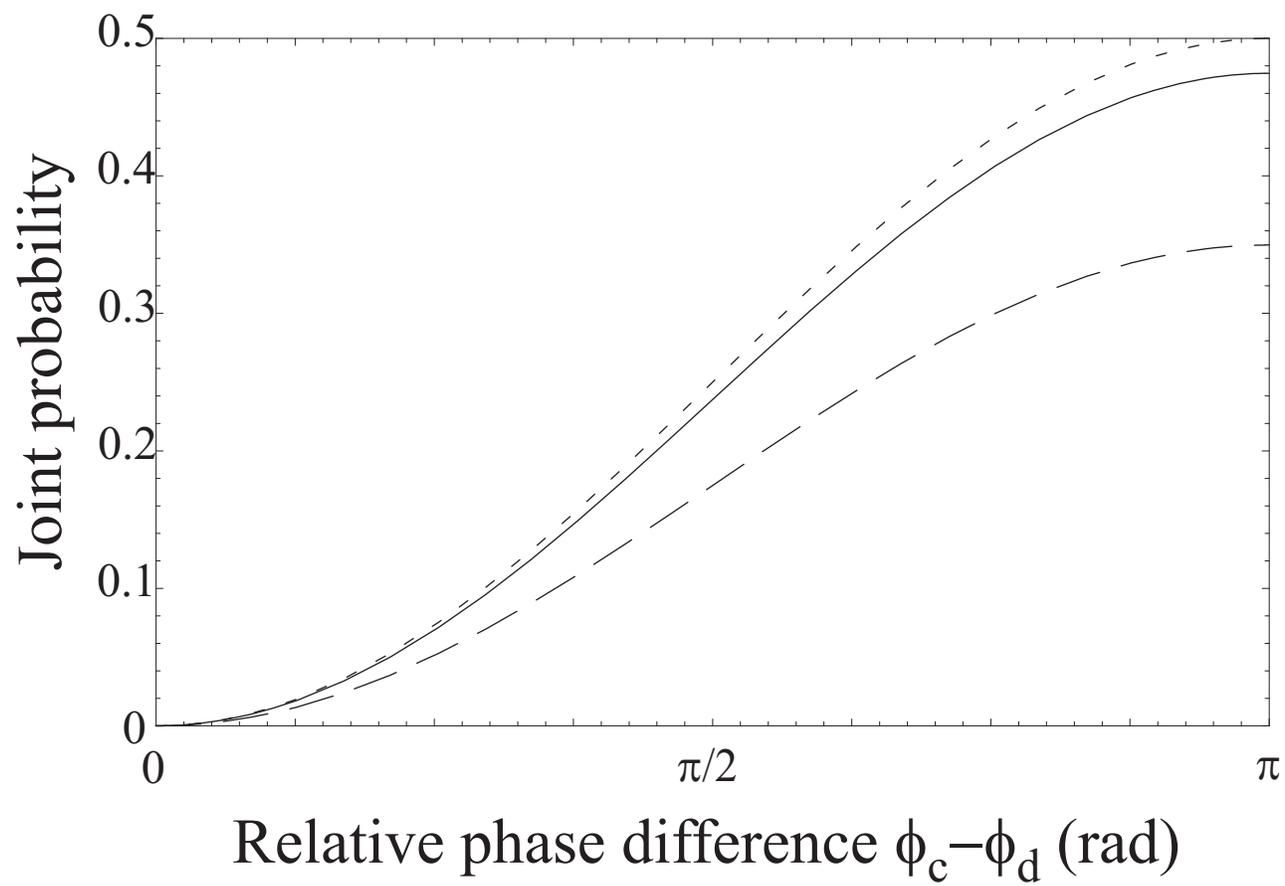
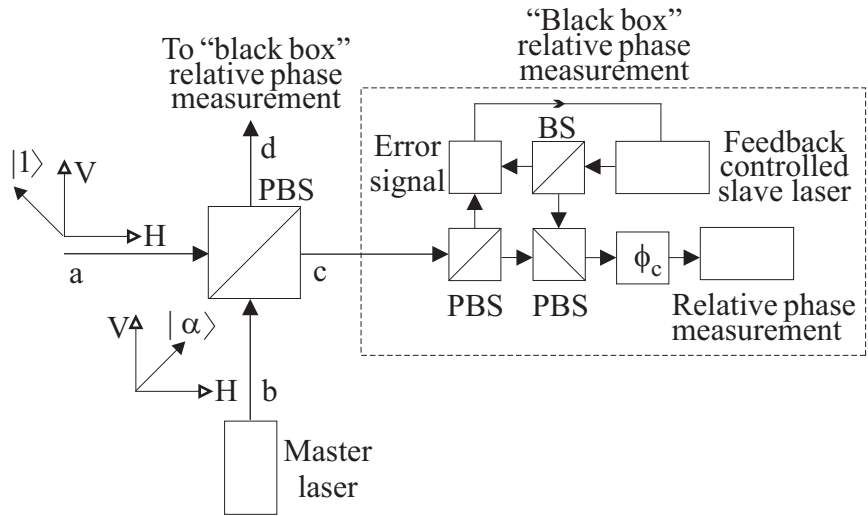
\end{document}